\documentclass[preprint,eqsecnum,floatfix]{revtex4}
\usepackage{graphicx}
\usepackage{epstopdf}
\newcommand{\dt}{{\ep}}
\newcommand{\e}{{\rm e}}
\newcommand{\ep}{\epsilon}
\newcommand{\al}{\alpha}
\newcommand{\ld}{\lambda}

\newcommand{\ga}{\gamma}
\newcommand{\ka}{\kappa}

\newcommand{\za}{\zeta}
\newcommand{\lm}{\lambda}
\newcommand{\pa}{{\partial}}

\newcommand{\bea}{\begin{eqnarray}}
\newcommand{\eea}{\end{eqnarray}}
\newcommand{\be}{\begin{equation}}
\newcommand{\ee}{\end{equation}}
\newcommand{\ba}{\begin{eqnarray}}
\newcommand{\ea}{\end{eqnarray}}

\newcommand{\nn}{\nonumber}
\newcommand{\la}{\label} 
\newcommand{\w}{\omega} 
\newcommand{\hT}{\hat{T}}
\newcommand{\hG}{\hat{G}}
\newcommand{\hV}{\hat{V}}

\begin{document}
\title{Analytical evaluations of the Path Integral Monte Carlo thermodynamic and
Hamiltonian energies\\ for the harmonic oscillator}
	
\author{Siu A. Chin}
\email{chin@physics.tamu.edu.}
	
\affiliation{Department of Physics and Astronomy, 
		Texas A\&M University, College Station, TX 77843, USA}
	
\begin{abstract}
By use of the recently derived {\it universal }
discrete imaginary-time propagator of the harmonic oscillator,
both thermodynamic and Hamiltonian energies can be given analytically,
and evaluated numerically at each imaginary time step, for $any$ short-time propagator.
This work shows that, using only currently known short-time propagators, the Hamiltonian 
energy can be optimized to the twelfth-order, converging to the ground state energy of
the harmonic oscillator in as few as three beads.
This study makes it absolutely clear that the widely used second-order primitive
approximation propagator, when used in computing the thermodynamic energy, 
converges extremely slowly with increasing number of beads.

\end{abstract}

\maketitle

\section {Introduction}

The second-order {\it primitive approximation} (PA) propagator has been widely used in
Path Integral Monte Carlo (PIMC) since its inception\cite{bar79,sch81,her82}.
However, even after the introduction of the fourth-order trace Takahashi-Imada (TI) propagator\cite{tak84}, 
it was not realized that the PA propagator's convergence was so very poor until recently when 
its energies were compared with those from truly fourth-order propagators\cite{jan01,whi07,sak09,kam16,lin18,wan22}.
Moreover, the wide-spread use of the PA propagator in the past has left a lasting, 
but misleading impression that PIMC generally requires hundreds of short-time propagators
in order to extract the ground state energy. Because of this, when the PA propagator 
is applied to fermion problems, the resulting sign problem is intractable and  
was later misinterpreted as being fundamental. This myth was only dispelled when the sign problem in 
quantum dots was greatly ameliorated when fourth-order propagators\cite{chin15} were used to compute 
the Hamiltonian energy\cite{whi07,chin15,rot10,chin20}. 

The reason why it has taken so long to assess the effectiveness of various short-time
propagators is the lack of a simple model for doing PIMC, where everything can be computed analytically. 
There is no analogous ``hydrogen atom" model for doing PIMC.
Even for the simple harmonic oscillator, where the discreet path integral
is known analytically\cite{sch81,kam16,wan22}, Sakkos, Casulleras and Boronat\cite{sak09} still have to
compute the thermodynamic energy by direct simulation. This is because
when the imaginary-time harmonic oscillator path integral is integrated directly\cite{sch81,kam16,wan22}, 
there is no simple way of changing the short-time propagator. Each short-time propagator give rises
to a different tri-diagonal matrix, which must be individually diagonalized to obtain the 
partition function for computing physical observables\cite{kam16,wan22}.
There is therefore a general lack of analytical tools for analyzing PIMC. 

However, recently, the ``hydrogen atom" model for doing PIMC has been found, which is
the {\it universal}, discrete imaginary time propagator for the harmonic oscillator\cite{chin22}.
This universal propagator was derived by completely abandoning the direct integration approach
with its obscuring tri-diagonal matrix. 
Instead, it was derived by contracting two free propagators into one in the presence of
the harmonic interaction\cite{chin22}. This then contracts {\it all} short-time propagators into a
standard form, like that of PA, but with different coefficient functions. Each short-time propagator can then 
be further contracted $N$ times to arrive at the discrete propagator at the $N^{th}$ imaginary time-step. 
The discrete propagator is universal in that it has the same functional form for all short-time propagators. 
Each short-time propagator only modifies the {\it portal} parameter, the argument
of the discrete propagator's hyperbolic functions. This clear separation between the discrete
propagator at any imaginary time step, and its dependence on any short-time propagator, makes it possible
to obtain physical observables in closed forms for all short-time propagators at once.

In this work, using the universal propagator, we present an analytical 
study of the thermodynamic and Hamiltonian energies at discrete imaginary time steps
in unprecedented details. This study is intended to be an in-depth reference for future PIMC 
simulations on the harmonic oscillator, especially when testing fourth and higher order short-time propagators. 
For example, we analytically derived here all of Sakkos {\it et al.}'s
discrete PIMC harmonic oscillator energies\cite{sak09}. We also state explicitly in this work, 
the {\it two fundamental convergence formulas} for the thermodynamic energy (proved in Ref.\cite{chin22}) 
and the Hamiltonian energy (derived here).
The Hamiltonian energy\cite{whi07,chin15,rot10,chin20}, 
is more complicated to evaluate and to optimize, but can converge at {\it twice} the
order of the thermodynamic energy. Knowing the convergence of the Hamiltonian energy
allows this work to derive a twelfth-order algorithm for its computation, capable 
of converging to the harmonic oscillator's ground state energy in as few as three beads.
These high order methods, when generalized to non-harmonic interactions, can greatly improve 
the efficiency of future PIMC, especially when solving fermion problems.

In Sect.\ref{udp}, we summarize salient features of the universal discrete propagator for the harmonic oscillator\cite{chin22}. The analytical form of the thermodynamics energy is given in Sect.\ref{tee}
together with a detailed comparison with Sakkos {\it et al.}'s\cite{sak09} PIMC data.
General fourth and higher order algorithms are described in Sect.\ref{stp}.
The Hamiltonian energy is given in Sect.\ref{hee} and its optimization, which differs from that
of the thermodynamic energy, is described in Sect.\ref{heop}. Here, we derive the condition 
under which the Hamiltonian energy can converge at twice the order of the thermodynamic energy
and showcase a twelfth-order propagator. The evaluation of the Hamiltonian energy is especially
important for solving fermion problems\cite{chin20} and for doing Path Integral Monte Carlo Ground State (PIMCGS)
calculations\cite{chin20,sar00,yan17}.
Conclusions are drawn in Sect.\ref{con}.

\section {The universal discrete propagator}
\la{udp}

To make this work self-contained, we summarize essential features of the universal discrete propagator 
derived in Ref.\onlinecite{chin22}.
In harmonic units, where lengths are in unit of the harmonic length $a=\sqrt{\hbar/(m\w)}$,
energy in terms of the harmonic energy $e_0=\hbar\w$ and with dimensionless imaginary time $\tau=e_0/(k_B T)$,
the dimensionless Hamiltonian operator for the the one-dimensional harmonic oscillator
\be
\hat H=
-\frac12\frac{d^2}{d { x}^2}
+\frac12 x^2\equiv\hat T+\hat V,
\la{hoh}
\ee
obey the operator identity:
\be
\e^{ -a\hT}\e^{ -b\hV}\e^{-c\hT}=\e^{ -\nu\hV}\e^{-\ka\hT}\e^{ -\mu\hV},
\la{opid}
\ee
where two free-propagators $\e^{-a\hT}$ and $\e^{-c\hT}$ can be contracted into one with
\be
\ka=a+abc+c,\quad\nu=\frac{bc}{\ka},\quad\mu=\frac{ba}{\ka}.
\la{kauv}
\ee
This means that any left-right symmetric approximate short-time operator of the form 
\be
\hG_1(\dt)=\prod_{i}\e^{ -a_i\dt\hT}\e^{ -b_i\dt\hV},
\ee 
can be contracted down to a single $\hT$-operator, palindromic form
\be
\hG_1(\dt)=\e^{ -\mu_1\hV}\e^{ -\ka_1\hT}\e^{ -\mu_1\hV},
\la{sho}
\ee 
where $\mu_1$ and $\ka_1$ are functions of $a_i$, $b_i$ and $\dt$, and whose matrix element 
\be
G_1(x^\prime,x,\ep)=\langle x^\prime|\hG_1(\dt)|x\rangle=\frac1{\sqrt{2\pi \ka_1(\ep)}}\e^{-\mu_1(\ep)\frac12 {x^\prime}^2}
\e^{-\frac1{2\ka_1(\ep)}(x^\prime-x)^2}
\e^{-\mu_1(\ep)\frac12 x^2},
\la{sp}
\ee
is the short-time propagator. 

For example, consider 
the following left-right symmetric operator with parameter $\al$,
\be
\hG_1(\dt)=
\e^{-\ep\hat V/2-\al\ep^3[\hat V,[\hat T,\hat V]]}
\e^{-\ep \hat T }
\e^{-\ep\hat V/2-\al\ep^3[\hat V,[\hat T,\hat V]]}.
\label{tiop}
\ee
For the harmonic oscillator, $[\hat V,[\hat T,\hat V]]=[V^\prime(x)]^2=x^2$, 
and therefore
\be 
\ka_1(\dt)=\dt\quad{\rm and}\quad \mu_1(\dt)=\dt/2+2\al\ep^3.
\la{paia}
\ee
Since (\ref{tiop}) is left-right symmetric it must be at least a second-order algorithm
for all $\al>0$. The choice of $\al=0$ corresponds to 
the original PA propagator\cite{bar79,sch81,her82}.
As will be shown in the next Section, as one
increases $\al$, one decreases its thermodynamic energy's second-order
error. When $\al$ reaches $\al=1/48$, the second-order error vanishes and the algorithm
becomes the fourth-order Takahashi–Imada (TI) propagator\cite{tak84}. This
tuning of $\al$ to increase the convergence order of the thermodynamic energy by two is
a also possible with later higher order propagators.

Since the transition between operator (\ref{sho}) and its matrix element (\ref{sp}) 
is trivial, we will refer to both forms as the ``short-time propagator". 

As also shown in Ref.\onlinecite{chin22}, the contraction of $N$ short-time propagator
$\hG_1(\dt)$, gives the discrete propagator at $\tau=N\ep$
\be
\hG_N=[\hG_1(\dt)]^N=\e^{ -\mu_N\hV}\e^{ -\ka_N\hT}\e^{ -\mu_N\hV},
\la{gn}
\ee
with {\it universal} coefficients
\ba
&&\za_N=\cosh(Nu),\la{hza} \\
&&\ka_N=\frac1{\ga}\sinh(Nu),\la{kka}\\
&&\mu_N=\frac{\za_N(Nu)-1}{\ka_N(Nu)}=\ga\tanh(Nu/2).
\la{mma}
\ea
Given $\ka_1$ and $\mu_1$, the above equations at $N=1$ define all quantities needed for $N>1$.
For example, $\za_1=1+\ka_1\mu_1$ is given by (\ref{mma}), from which
the {\it portal} parameter $u$ is defined from (\ref{hza})
\be
u=\cosh^{-1}(\za_1)=\ln\left(\za_1(\ep)+\sqrt{\za_1^2(\ep)-1}\right),
\la{udef}
\ee
and where $\ga$, once defined from $N=1$, remains true for all $N$:
\be
\ga=\frac{\sinh(u)}{\ka_1}=\frac{\sqrt{\za_1^2-1}}{\ka_1}=\frac{\sqrt{\za_N^2-1}}{\ka_N}.
\la{gai}
\ee
The discrete propagator is a universal function of $u$ with dependence on any short-time propagator 
only through the portal parameter $u$ via (\ref{udef}).

\section {The Thermodynamic energy}
\la{tee}

\begin{figure}[hbt]
	\includegraphics[width=0.48\linewidth]{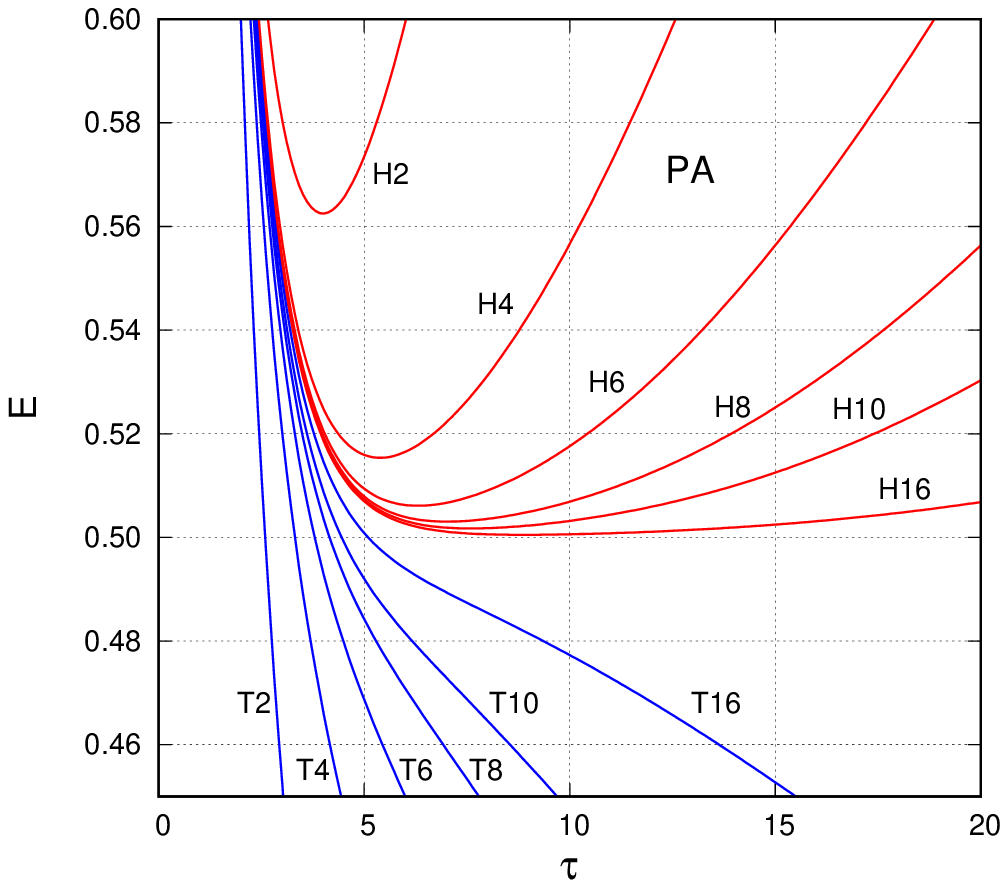}
	\includegraphics[width=0.48\linewidth]{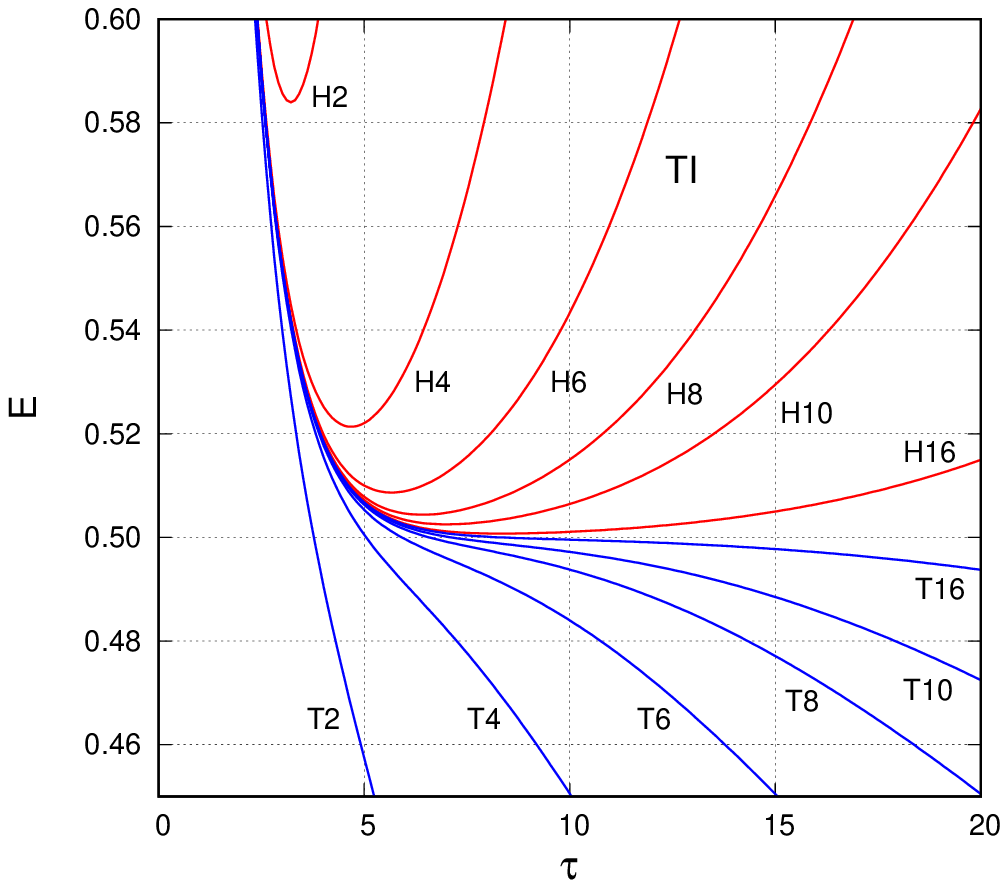}
	
	\includegraphics[width=0.48\linewidth]{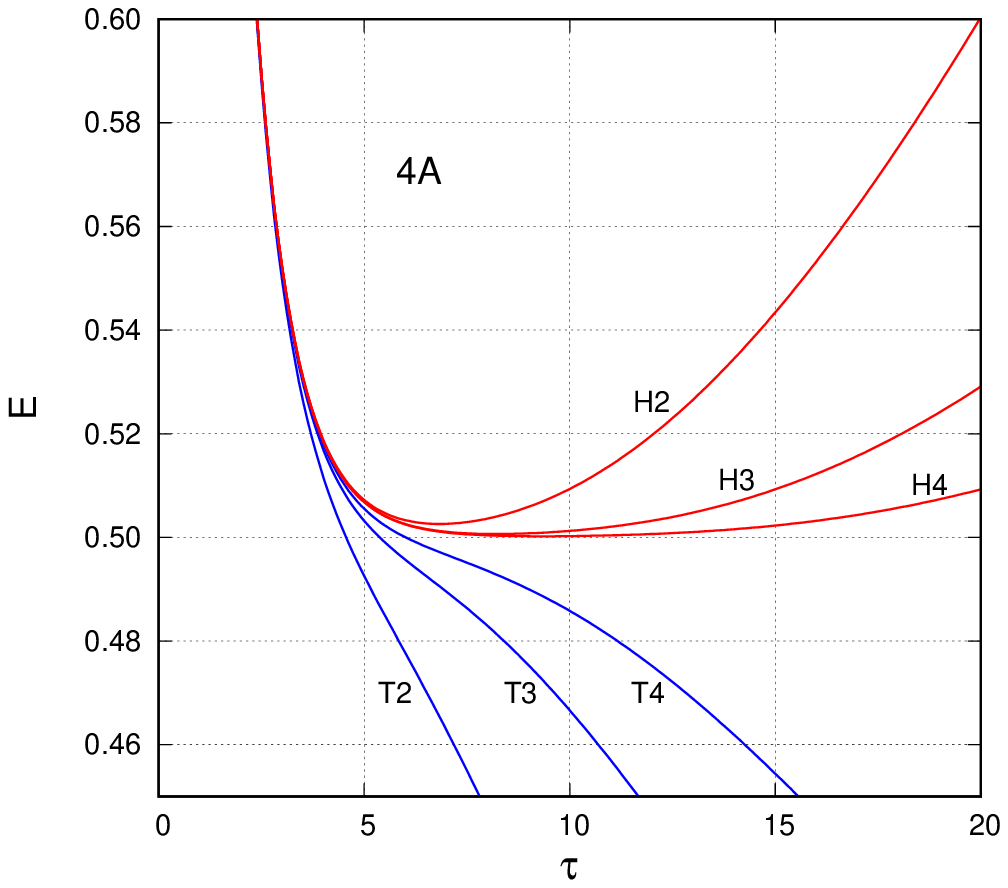}
	\includegraphics[width=0.48\linewidth]{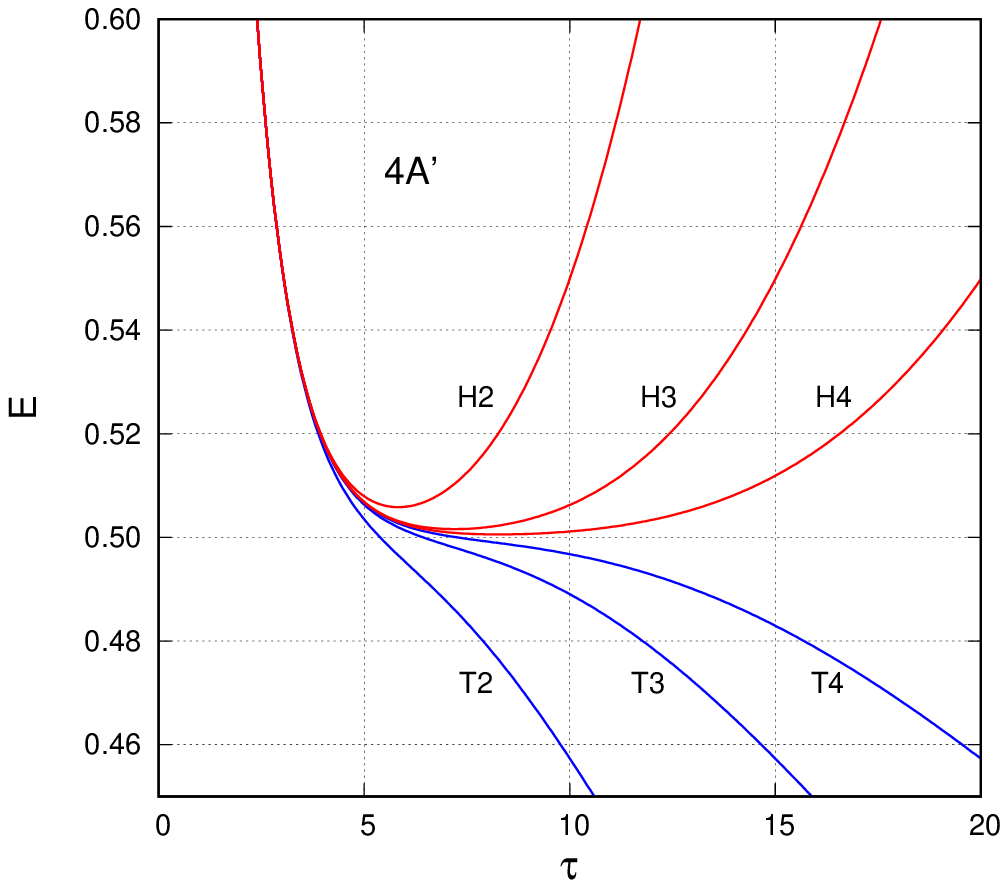}
	
	\includegraphics[width=0.48\linewidth]{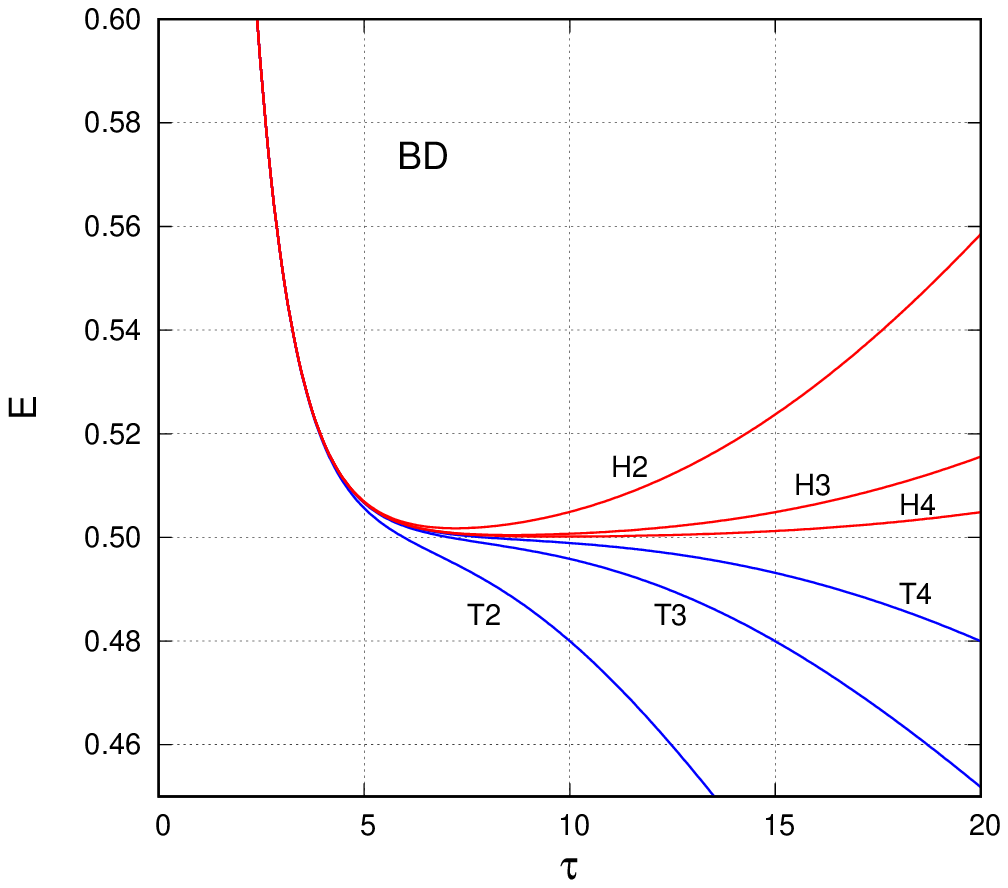}
		\includegraphics[width=0.48\linewidth]{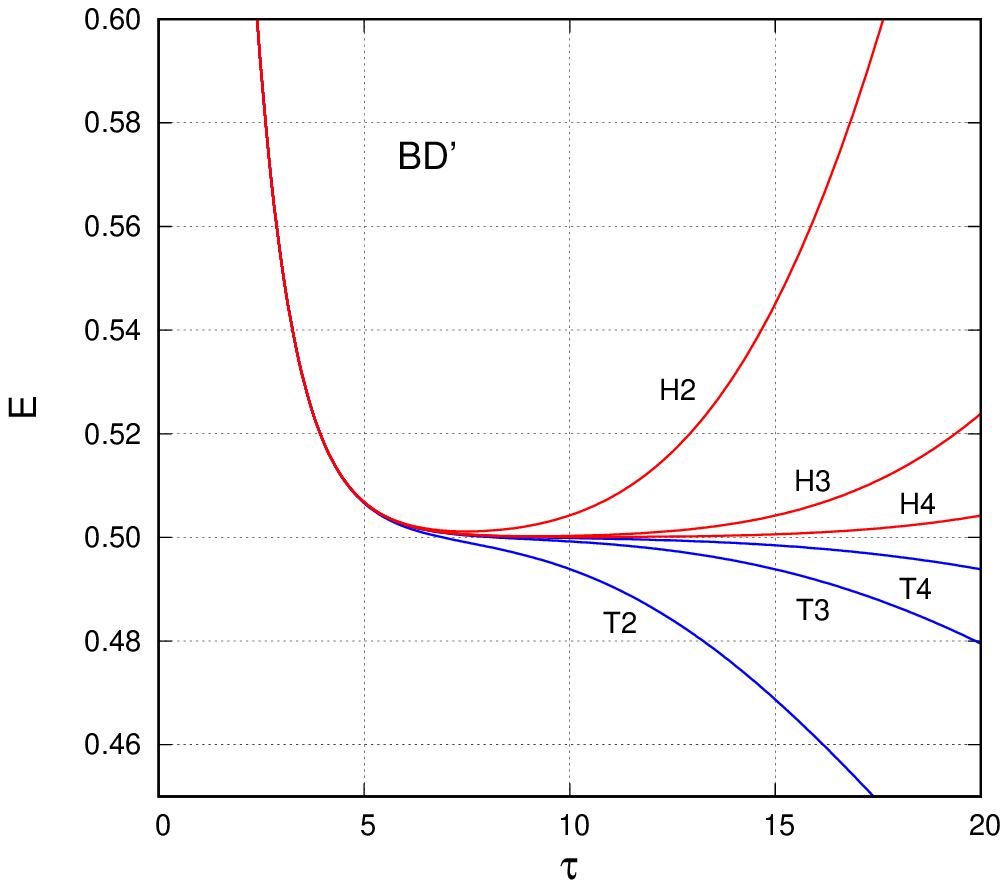}
	\caption{ (color online) 
		The $N$-bead thermodynamic and Hamiltonian energies, denoted by T$N$ and H$N$ respectively,
		are plotted as a function of the imaginary time $\tau=N\ep$, for  
		short-time propagators PA, TI, 4A, 4A$^\prime$, BD and BD$^\prime$. See text for details.	}
	\la{engpa}
\end{figure}

From Ref.\onlinecite{chin22}, the $N$-bead thermodynamic energy is
\ba
E_N^T(\ep)&=&-\frac1{Z_N(\ep)}\frac{d Z_N(\ep)}{d \dt},
\la{zn}
\ea
where the discrete partition function is given by.
\ba
Z_N(\ep)&=&\int dx G_N(x,x,\dt)=\frac1{\sqrt{2\pi \ka_N}}
\sqrt{\frac{\pi}{\mu_N}},\nn\\
&=&\frac1{\sqrt{2(\za_N-1)}}=\frac1{2 \sinh(Nu/2)}.
\la{trce}
\ea
The discrete thermodynamics energy (\ref{zn}) is therefore
\be
E^T_N(\ep)=\rho_T E_N(\ep),
\la{znf}
\ee
where the prefactor is given by (\ref{udef})
\be
\rho_T=\frac{du}{d\ep}=\frac1{\sqrt{\za_1^2-1}}\frac{d\za_1}{d\ep}
\la{dude}
\ee
and where 
\be
E_N(\ep)=\frac12\coth(Nu/2)=\frac12\frac{\ga}{\mu_N}.
\la{enu}
\ee
is the {\it universal discrete energy}, same for all short-time
propagators. Since $u$ is only a function of $\za_1$, the discrete thermodynamics energy
(\ref{znf}) is solely a function of $\za_1$.
However, for later comparison with the Hamiltonian energy, one can also write the prefactor
alternatively as
\be
\rho_T=\frac{\ka_1}{\sqrt{\za_1^2-1}}
\left (\frac1{\ka_1}\frac{d\za_1}{d\ep}\right)\equiv\frac1{\ga}\ld,
\ee
in term of the old $\ga$ and a newly defined $\ld$.

For PA, $\al=0$ in (\ref{paia}) gives $\za_1=1+\ep^2/2$ and
\be
E^T_N(\ep)=\frac{1}{\sqrt{1+\ep^2/4}}E_N(\ep).
\la{thpa}
\ee
For TI,  $\al=1/48$ in (\ref{paia}) gives $\za_1=1+\ep^2/2+\ep^4/24$ and
\be
E^T_N(\ep)
=\frac{1+\ep^2/6}{\sqrt{1+\ep^2/3+\ep^4/24+\ep^6/576}}E_N(\ep).
\la{thti}
\ee
These PA and TI thermodynamic energies, (\ref{thpa}), (\ref{thti}), are plotted as a function
of $\tau$ for $N=2$ to 16 with $\ep=\tau/N$ in the top two graphs of Fig.\ref{engpa}.
At these small values of $N$, the thermodynamic energies showed no
convergence whatsoever for the PA propagator.  This has left a lasting, but misleading impression 
that hundreds of beads are needed for energies in PIMC to converge (see also Table I).
This is clearly not the case when the Hamiltonian energy is computed. For TI, the 
thermodynamic energies are markedly improved, because it is fourth-order. However, its
Hamiltonian energies have gotten worse, as we will explain in the Sect.\ref{hee}.

\begin{table}
	\caption{Columns 1 to 4 compares $N$-bead PIMC thermodynamic energies of Sakkos {\it et al.}\cite{sak09} at 
		$\tau=5$ with this work's analytical PA results (\ref{thpa}) and TI results (\ref{thti}).}
	\begin{center}
		\begin{tabular}{c r r r r r r  }
			\colrule
			$N$ &\ \ Ref.\onlinecite{sak09}'s\ PA&\ \ PA \ \ \ \ \  &\ \ Ref.\onlinecite{sak09}'s\ TI\ \ \ &TI\ \ \ \ \ \ \ \\ 
			\colrule
			2      &0.30755& \ \ \ 0.32195122   &   0.44702 \ \ \ & 0.45746346  \\
			4      &0.43162& \ \ \ 0.43161837   &   0.50053 \ \ \ & 0.50053259   \\ 
			8      &0.48424& \ \ \ 0.84243957   &   0.50630 \ \ \ & 0.50629474   \\        
			16     &0.50085& \ \ \ 0.50084554   &   0.50675 \ \ \ & 0.50675121  \\         
			32     &0.50528& \ \ \ 0.50527874   &   0.50678 \ \ \ & 0.50678160   \\        
			64     &0.50641& \ \ \ 0.50640613   &   0.50678 \ \ \ & 0.50678353  \\         
			128    &0.50669& \ \ \ 0.50668919   &   \ \         & 0.50678365   \\        
			256    &0.50676& \ \ \ 0.50676003   &   \ \         & 0.50678365   \\      
			512    &0.50678& \ \ \ 0.50677775   &  \ \          & 0.50678366   \\        
			1024   &       & \ \ \ 0.50678218  &   \ \          & 0.50678366     \\     
			$\infty$ & \ \ \   & \ \ \ \ \ 0.50678366 & \ \  &  \\     
			\colrule
		\end{tabular}
	\end{center}
	\label{tab1}
\end{table}

In Table \ref{tab1}, our numerically evaluated thermodynamic
energies for PA and TI 
are compared  with those of Sakkos {\it et al.}'s PIMC results\cite{sak09} at $\tau=5$ as a function of $N$.
This table concretely illustrates the slow convergence of 
PA's thermodynamic energy. 
If Ref.\onlinecite{sak09}'s 
final PA value of 0.50678 is regarded as the converged result, then the use of
PA requires $N=512$. As shown in Table II, the same value is obtained by the sixth-order BD$^*$ algorithm
at only $N=3$, when computing the Hamiltonian energy.

For small $N$, one can obtain the thermodynamic energy analytically, not just numerically from
(\ref{thpa}) and (\ref{thti}).
Instead of (\ref{udef}), one can also write
\be
\za_1=1+2\sinh^2(u/2)\quad\rightarrow\quad\sqrt{\frac{\za_1-1}2}=\sinh(u/2),
\la{sinh2}
\ee
so that
\be
\e^{u/2}=\sqrt{\frac{\za_1(\ep)+1}2}+\sqrt{\frac{\za_1(\ep)-1}2}\equiv g.
\la{u2}
\ee
and
\be
g^{-1}=\sqrt{\frac{\za_1(\ep)+1}2}-\sqrt{\frac{\za_1(\ep)-1}2}.
\ee
The discrete thermodynamics energy (\ref{enu}) can then be expressed as
\ba
E_N
&=&\frac12\frac{(g^N+g^{-N})}{(g^N-g^{-N})}.
\la{ethuni}
\ea
Combining the above with (\ref{znf}) then gives
\ba
E^T_1&=&\frac12 \frac1{\za_1-1}\frac{d\za_1}{d\ep},\nn\\
E^T_2&=&\frac12 \frac{\za_1}{\za_1^2-1}\frac{d\za_1}{d\ep},\nn\\
E^T_4&=&\frac12 \frac{\za^2_1-1/2}{\za_1(\za_1^2-1)}\frac{d\za_1}{d\ep}.
\ea
For PA, $\za_1=1+\ep^2/2$, $d\za_1/d\ep=\ep$ and $\ep=\tau/N$ gives
\ba
E^T_1(\tau)&=&\frac1{\tau},\nn\\
E^T_2(\tau)&=&\frac1{\tau}\frac{(1+\frac{\tau^2}8)}{(1+\frac{\tau^2}{16})},\nn\\
E^T_4(\tau)&=&\frac1{\tau}\frac{(1+\frac{\tau^2}8+\frac{\tau^4}{512})}
{(1+\frac{3\tau^2}{64}+\frac{\tau^4}{2048})}.
\ea
At $\tau=5$, 
$E^T_2(5)$=$\frac{66}{205}$=0.32195122 and
$E^T_4(5)$=$\frac{10948}{25365}$=0.43161837
are as shown in Table I.
It is not clear why Ref.\onlinecite{sak09}'s empirical findings, otherwise
in excellent agreement with our numerical results, differs from our analytical two-bead energy $E^T_2(5)$.

For TI, $\za_1=1+\ep^2/2+\ep^4/24$, $d\za_1/d\ep=\ep+\ep^3/6$ and $\ep=\tau/N$ gives
\ba
E^T_1(\tau)&=&\frac1{\tau}\frac{(1+\frac{\tau^2}6)}{(1+\frac{\tau^2}{12})},\nn\\
E^T_2(\tau)&=&\frac1{\tau}\frac{(1+\frac{\tau^2}{24})(1+\frac{\tau^2}8+\frac{\tau^4}{384})}
{(1+\frac{\tau^2}{12}+\frac{\tau^4}{384}+\frac{\tau^6}{36864})},\nn\\
E^T_4(\tau)&=&\frac1{\tau}\frac{(1+\frac{\tau^2}{96})(1+\frac{\tau^2}8
	+\frac{\tau^4}{384}+\frac{\tau^6}{49152}+\frac{\tau^8}{18874368})}
{(1+\frac{5\tau^2}{96}+\frac{\tau^4}{1024}+\frac{7\tau^6}{786432}
	+\frac{\tau^8}{25165824}+\frac{\tau^{10}}{14495514624})}.
\ea
At $\tau=5$, 
$E^T_2(5)$=$\frac{432964}{946445}$=0.45746346 and
$E^T_4(5)$=$\frac{111288436424}{222340041245}$=0.50053259 are also shown in Table I. Again, 
$E^T_2(5)$ above disagrees with Sakkos {\it et al.}'s\cite{sak09} result.

In the convergence limit of fixed $\tau$ but with $N\rightarrow \infty$ and $\ep=\tau/N \rightarrow 0$,
$u\rightarrow\ep$ and
\be
E_N(\ep)\rightarrow E(\tau)=(1/2)\coth(\tau/2),
\ee
(\ref{thpa}) for PA gives
\be
E^T_N\rightarrow
(1-\frac1{8}\ep^2+\frac3{128}\ep^4+\cdots)E(\tau),
\la{palim}
\ee
verifying that its convergence is second order in $\ep$ from below. For TI, 
(\ref{thti}) gives
\be
E^T_N\rightarrow
(1-\frac1{144}\ep^4+\cdots)E(\tau),
\la{tilim}
\ee
and its convergence is fourth-order from below. 

These results are special cases of the
{\it fundamental convergence formula for the thermodynamic energy} (\ref{econg}) below,
first proved in Ref.\onlinecite{chin22}. If a short-time propagator's $\za_1$ is correct up to order $2n$, 
but has error at order $2n+2$ with coefficient $\delta_{2n+2}$,
\be
\za_1=1+\frac1{2!}\ep^2+\cdots +\frac1{(2n)!}\ep^{2n}+\frac{\delta_{2n+2}}{(2n+2)!}\ep^{2n+2},
\ee
then the portal parameter must converge with error $\ep^{2n}$
\be
u=\ep \left (1-\frac{(1-\delta_{2n+2})}{(2n+2)!}\ep^{2n} +O(\ep^{2n+2}) \right ),
\la{ufin}
\ee
and the thermodynamic energy converges to the same order as
\be
E^T_N(\tau)
\rightarrow \left (1-(2n+1)\frac{(1-\delta_{2n+2})}{(2n+2)!}\ep^{2n}+O(\ep^{2n+2})\cdots\right )E(\tau),
\la{econg}
\ee
provided that the error in $E_N(\tau)$ can be ignored at sufficiently 
large value of $\tau$, on the order of $\tau\gtrsim 10$. (If $\tau$ is not sufficiently large,
then one must include the discrete error in $E_N(\tau)$, see Ref.\onlinecite{chin22}.)

For the PA-TI pair, (\ref{paia}) gives $\za_1=1+\ep^2/2+2\al\ep^4$, $n=1$, with $\delta_4=2\al 4!=48\al$.
At $\al=0$, (\ref{econg}) reproduces PA's leading error term in (\ref{palim}). As $\al$ increases,
the second-order error in (\ref{econg}) decreases and vanishes at $\al=1/48$, yielding the
TI algorithm with $n=2$, $\delta_6=0$ and error in (\ref{econg}) now matches (\ref{tilim}). 

The PA-TI pair is a single algorithm with one free parameter $\al$. As we have seen,
$\al=1/48$ optimizes the thermodynamic energy to the fourth-order. As we will see in Sect.\ref{heop},
$\al=0$ optimizes the Hamiltonian energy to the fourth-order. This optimization pattern is
exactly replicated by the bona fide fourth-order  
short-time propagator\cite{chin02} 4A, also with one free parameter $\al$:
\be
\hG_1= 
\e^{-\frac16\ep\hV_0}
\e^{-\frac{1}2\ep \hat T}
\e^{-\frac23\ep\hV_1}
\e^{-\frac{1}2\ep \hat T}
\e^{-\frac16\ep\hV_0},
\ee
\ba
&&\qquad\frac16\hV_0=\frac16 \hV+\frac{\al}{2}\frac{\epsilon^2}{72}[\hV,[\hT,\hV]]\nn\\
&&\qquad\frac23\hV_1=\frac23 \hV+(1-\al)\frac{\epsilon^2}{72}[\hV,[\hT,\hV]].
\label{vtbd}
\ea
Contracting the two $\hT$ operators using (\ref{opid}) gives
\ba
\ka_1&=&a(2+ab)=\ep(1+\frac{\ep}4 b)=\ep+\frac1{3!}\ep^3+\frac{1-\al}{144}\ep^5\la{ka4a}\\
\za_1&=&1+\ka_1\left [\frac{ba}{\ka_1}+\ep(\frac16+\al\frac{\ep^2}{72})\right ]\nn\\
&=&1+\frac{\ep^2}{2}+\frac{\ep^4}{4!}+\frac{1+\alpha}{864}\ep^6+\frac{\alpha(1-\alpha)}{10368}\ep^8.
\la{za4a}
\ea
The case of $\al=0$ is the original 4A algorithm with a fourth-order
thermodynamic energy. 
The choice of $\alpha=1/5$, denoted as algorithm 4A$^\prime$, 
would force $(1+\al)/864=1/6!$ resulting in a sixth-order thermodynamic energy according to (\ref{econg}). 
This is then the exact analog of PA and TI, just two orders higher. 
Their thermodynamic energies for $N=2,3,4$ are plotted in the middle row of
Fig.\ref{engpa}. At these low values of $N$, despite the improvement with 4A$^\prime$,
the thermodynamic energies showed no convergence, in contrast
to the dramatically improved Hamiltonian energies of 4A, to be discussed in Sect.\ref{hee}.

While TI is a general trace fourth-order algorithm for all interactions, propagator 4A$'$ is only
sixth-order for the harmonic oscillator. 
This is because for PA, tuning $\al$ can set coefficients of {\it two} second-order error commutators 
in the algorithm's Hamiltonian equal, resulting in TI. For 4A, there are
generally four fourth-order error commutators, and their coefficients must be set equal by pairs 
to produce a trace sixth-order algorithm. This would require at least two free parameters. 
In the special case of the harmonic oscillator, one pair of commutators
vanishes identically, therefore, only one parameter is needed, allowing 4A$'$ to be sixth order. 
For a detailed discussion, see Ref.\onlinecite{chin07}. 

For non-harmonic interactions, one can fine tune $\al$ to reduce the fourth-order error in 4A as much as possible,
but there is no guarantee that it can yield a sixth-order algorithm.

\section {Higher order short time propagators}
\la{stp}

For both PA-TI and 4A-4A$'$, the parameter $\al$ can only tune the thermodynamic energy two orders higher.
At the three-$\hT$ level, one has the
BDA family of fourth-order short-time propagator\cite{chin02}
given by
\be
	{\cal T}_{BDA}=
	{\rm e}^{-v_0\epsilon \hV_0}
	{\rm e}^{-t_1\epsilon \hT} 
	{\rm e}^{-v_1\epsilon \hV_1}
	{\rm e}^{-t_2\epsilon \hT} 
	{\rm e}^{-v_1\epsilon \hV_1}
	{\rm e}^{-t_1\epsilon \hT}
	{\rm e}^{-v_0\epsilon \hV_0},
	\label{algbda}
\ee
\be
t_2=1-2t_1,\quad
v_1=\frac1{12t_1(1-t_1)},\quad 
v_0=\frac12-v_1,
\ee
\ba
v_0\hV_0&=&v_0\hV+\al u_0\epsilon^2[\hV,[\hT,\hV]],\nn\\
v_1\hV_1&=&v_1\hV+(1-\al)u_0\epsilon^2[\hV,[\hT,\hV]],\nn\\
u_0&=&{1\over 48}\biggl[{1\over{6t_1(1-t_1)^2}}-1\biggr],
\label{vtbd}
\ea
with two free parameters are ${1\over 2}(1-{1\over{\sqrt 3}})\leq t_1\leq {1\over 2}$ and $0\leq\al\leq 1$.
The contraction of (\ref{algbda}) now gives
\ba
\ka_1&=&\ep+\frac{\ep^3}{3!}+\delta_5'(\al,t_1)\frac{\ep^5}{5!}+\delta_7'(\al,t_1)\frac{\ep^7}{7!}+\cdots\nn\\
\za_1
&=&1+\frac{\ep^2}{2!}+\frac{\ep^4}{4!}+\delta_6(\al,t_1)\frac{\ep^6}{6!}
+\delta_8(\al,t_1)\frac{\ep^8}{8!}+\cdots.
\ea
As shown in Ref.\onlinecite{chin22}, one can always solve $\delta_6(\al,t_1)=1$
for $\al$ to yield a sixth-order thermodynamic energy algorithm, then solve for 
$t_1$ to maximize $\delta_8$. This yielded $t_1=0.27564$ and $\al=0.171438$,
giving $\delta_8=0.98967$, which is very close to an eighth-order algorithm ($\delta_8=1$). 
This optimized form of (\ref{algbda}) will be
designated as BD$'$ here and whose energies are shown on the bottom right of Fig.\ref{engpa}.
This algorithm's sixth-order thermodynamic energy error coefficient (\ref{econg}) is $\approx 30$ times smaller 
than that of 4A$'$ and the improvement is clearly visible.
Its Hamiltonian energy is even more dramatically lowered. As will be shown in Sect.\ref{heop},
with $\delta_5'=0.9647595$, its Hamiltonian energy, like that of 4A, remained only eighth-order, 
but with much smaller error coefficient than 4A.

Sakkos {\it et al.} \cite{sak09} have done extensive PIMC simulations
on their CA (``Chin action") algorithm, which is the ACB form of the 
propagator\cite{chin02} with four $\hT$-operators and distributed commutators\cite{san05}: 
\be
{\cal T}_{ACB}\equiv
{\rm e}^{ -t_0\ep\hT}
{\rm e}^{ -v_1\ep\hV_1}
{\rm e}^{ -t_1\ep\hT}
{\rm e}^{ -v_2\ep\hV_2}
{\rm e}^{ -t_1\ep\hT}
{\rm e}^{ -v_1\ep\hV_1}
{\rm e}^{- t_0\ep\hT}\, , \la{algacop}
\ee
where 
\be
t_1={1\over 2}-t_0 ,\quad
v_2=1-2v_1,\quad
v_1={1\over 6}{1\over{(1-2 t_0)^2}},
\la{acofac}
\ee
\ba
v_1\hV_1&=&v_1\hV+\frac{\al}2 u_0 \ep^2 [\hV,[\hT,\hV]]\, , \nn\\
v_2\hV_2&=&v_2\hV+(1-\al)u_0\ep^2[\hV,[\hT,\hV]]\, , \la{vtac2}\\
u_0&=&{1\over 12}\biggl[1-{1\over{1-2t_0}}+{1\over{6(1-2t_0)^3}}\biggr],\la{uo}
\ea
and where they use the parameter $a_1=\al/2$ instead. 
The two free parameters ($t_0$, $a_1$) algorithm which can be contracted\cite{chin22} to yield
\be
\za_1=1+\frac{\ep^2}{2}+\frac{\ep^4}{4!}
+\delta_6(a_1,t_0)\frac{\ep^6}{6!}
+\delta_8(a_1,t_0)\frac{\ep^8}{8!} +\cdots.
\la{wan}
\ee
Empirically, they found a set of ($t_0$, $a_1$) values for which
the algorithm yielded sixth-order thermodynamic energies. 
Their energies for two cases, CA$_1$($t_0=0.1430,a_1=0$) and CA$_2$($t_0=0.1215,a_1=0.33$),
are tabulated in Table \ref{tab2} and compared with this work's analytical result (\ref{znf}).
The agreements are excellent, with only slight differences in the fifth decimal
place in the $N=2,3$ cases of CA$_2$. The convergence of these two CA short-time propagators 
falls between 4A-4A$^\prime$ and $BD'$ and therefore was not shown in Fig.\ref{engpa}.
Their tabulated energies are also compared to that of BD$'$ in Table \ref{tab2}. At $N=4$, BD$'$
attained CA$_2$'s energy at $N=5$, CA$_1$'s energy at $N=6$, TI's energy at $N=32$ and PA's energy at $N=512$.

\begin{table}
	\caption{The first four columns compare $N$-bead PIMC thermodynamic energies of Sakkos {\it et al.}\cite{sak09} 
		at $\tau=5$ for CA$_1$ ($t_0=0.1430,a_1=0$) and CA$_2$ ($t_0=0.1215,a_1=0.33$) with analytical
		results given by (3.3). Column 5 gives the thermodynamics energy of algorithm BD and column 6
		gives the Hamiltonian energy of algorithm BD$^*$. }
	\begin{center}
		\begin{tabular}{c r r r r r r r r }
			\colrule
			$N$ & CA$_1$\ \ \ & Eq.(\ref{znf})\ \ \ \ \ & CA$_2$\ \ \ \ \ 
			    &  Eq.(\ref{znf})\ \ \ & BD$'$\ \ \ \ \ & HBD$^*$\ \ \ \ \\
			\colrule
 2      &0.50444& \ \ \ 0.50444339\ \ \  &   0.50643 \ \ \ & 0.50640167 &\ \  0.50660946 & \ \ 0.50679043 \\
 3      &0.50649& \ \ \ 0.50649346\ \ \  &   0.50675 \ \ \ & 0.50674038 &\ \  0.50676633 & \ \ 0.50678396 \\ 
 4      &0.50673& \ \ \ 0.50672790\ \ \  &   0.50677 \ \ \ & 0.50677521 &\ \  0.50678043 & \ \ 0.50678370 \\
 5      &0.50677& \ \ \ 0.50676964\ \ \  &   0.50678 \ \ \ & 0.50678137 &\ \  0.50678279 & \ \ 0.50678367 \\ 
 6      &0.50678& \ \ \ 0.50677953\ \ \  &   0.50678 \ \ \ & 0.50678289 &\ \  0.50678336 & \ \ 0.50678366 \\
 7      &0.50678& \ \ \ 0.50678235\ \ \  &           \ \ \ & 0.50678336 &\ \  0.50678354 & \ \ 0.50678366 \\  			
$\infty$ & \ \  & \ \ \ 0.50678366\ \ \ & \ \  &  & \\     
			\colrule
		\end{tabular}
	\end{center}
	\label{tab2}
\end{table}

\section {The Hamiltonian energy}
\la{hee}

The $N$-bead Hamiltonian energy is given by
\ba
E_N^{H}(\dt)&=&\lim_{x^\prime \rightarrow x}
\frac{\int dx H G_N(x,x^\prime,\dt)}{\int dx G_N(x,x,\dt) }.
\la{ehdef}
\ea
Denoting $G_N(x,x^\prime,\dt)=\e^{-U_N(x,x^\prime,\dt)}$ with
\be
U_N(x,x^\prime,\dt)=\frac12\ln(2\pi \ka_N)+
\frac12\mu_N(x^2+(x^\prime)^2)+\frac1{2\ka_N}(x-x^\prime)^2,
\ee
one has
\ba
\lim_{x^\prime \rightarrow x} H G_N(x,x^\prime,\dt) &=&\lim_{x^\prime \rightarrow x}
(-\frac12\frac{d^2}{dx^2}+\frac12 x^2)G_N(x,x^\prime,\dt)\nn\\
&=&\lim_{x^\prime \rightarrow x}\left(
\frac12 \frac{\partial^2 U_N}{\pa x^2}-\frac12 \left(\frac{\partial U_N}{\pa x}\right)^2+\frac12 x^2\right)
G_N(x,x^\prime,\dt)\nn\\
&=&\left(\frac12 (\mu_N+\frac1{\ka_N})+\frac12 (1-\mu_N^2) x^2\right)G_N(x,x,\dt)
\ea
and therefore the Hamiltonian energy is
\ba
E_N^{H}(\dt)&=&\frac12 (\mu_N+\frac1{\ka_N})+\frac12 (1-\mu_N^2)\frac{\int dx x^2 G_N(x,x,\dt)}{\int dx G_N(x,x,\dt) }\nn\\
&=&\frac12 (\mu_N+\frac1{\ka_N})+\frac12 (1-\mu_N^2)\frac1{2\mu_N}\nn\\
&=&\left[1+\mu_N(\mu_N+\frac2{\ka_N})\right]\frac1{4\mu_N} =(1+\ga^2)\frac1{4\mu_N}\nn\\
&=&\rho_H E_N(\ep),
\la{ehgen}
\ea
where one has recalled (\ref{enu}) and
\be
\rho_H=\frac1{2}(\ga+\frac1{\ga}).
\ee
The Hamiltonian energy has the same universal discrete energy $E_N(\ep)$ as the thermodynamic energy, but with a different prefactor $\rho_H$. Whereas $\rho_T$ depends solely on $\za_1$,
$\ga$ depends on both $\za_1$ {\it and} $\ka_1$. To optimize $E_N^{T}$, one only
has to match $\za_1(\ep)$ to $\cosh(\ep)$. As will be shown in the next Section, to optimize $E_N^{H}$, one must
optimize the relationship between $\za_1$ and $\ka_1$.

Comparing (\ref{ehgen}) to (\ref{znf}) with $\rho_T=\lm/\ga$ gives
the remarkable result, that at any $\tau=N\ep$, the {\it ratio of two discrete 
energies is solely determined by the short-time-propagator} via
\be
\frac{E_N^{H}(\tau)}{E_N^{T}(\tau)}=\frac1{2}\frac{1+\ga^2(\ep)}{\lm(\ep)}.
\ee
For PA, 
\be 
\lm=1\qquad{\rm and}\qquad\ga=\sqrt{1+\ep^2/4}
\la{gpa}
\ee
give
\be
E_N^{H}(\tau)=(1+\ep^2/8)E_N^{T}(\tau).
\ee
For TI, 
\be 
\lm=1+\ep^2/6\qquad{\rm and}\qquad\ga=\sqrt{1+\ep^2/3+\ep^4/24+\ep^6/576}
\la{gti}
\ee
give
\be
E_N^{H}(\tau)=\left(1+\frac{\ep^4/48+\ep^6/1152}{1+\ep^2/6}\right)E_N^{T}(\tau).
\ee
These are the Hamiltonian energies plotted in the upper two graphs in Fig.\ref{engpa},
which can also be directly evaluated from (\ref{ehgen}). They are variational and each
bead energy has a minimum which is much closed to the ground state energy than their
corresponding thermodynamic energy. 
In the convergence limit of fixed $\tau$ but $\ep\rightarrow 0$, the Hamiltonian energy for PA
approaches
\ba
E_N^{H}(\tau)&=&(1+\ep^2/8)E_N^{T}(\tau)=(1+\ep^2/8)(1-\frac1{8}\ep^2+\frac3{128}\ep^4+\cdots)E(\tau)\nn\\
&\rightarrow&(1+\frac1{128}\ep^4+\cdots)E(\tau),
\la{hpalm}
\ea 
which is fourth-order from above and twice the order of the thermodynamic energy. 
For TI, because it is only a trace fourth-order algorithm, despite the fact that its
thermodynamic energy is fourth-order, it remains only a second-order algorithm for non-trace calculations,
so that its Hamiltonian energy remains only fourth-order,
\ba
E_N^{H}(\tau)&=&\left(1+\frac{\ep^4/48+\ep^6/1152}{1+\ep^2/6}\right)(1-\frac1{144}\ep^4+\cdots)E(\tau)\nn\\
&\rightarrow& (1+\frac1{72}\ep^4+\cdots)E(\tau),
\la{htilm}
\ea
but with even a large error coefficient than PA. This is why in Fig.\ref{engpa}, TI's Hamiltonian energies are
higher than those of PA. A more comprehensive explanation will be given in the next Section. 

One can also obtain analytical forms for specific bead energies. For example, for PA,
the one and two-bead Hamiltonian energies are respectively,
\be
E_1^{H}(\tau)=(1+\ep^2/8)E_1^{T}(\tau)=(1+\tau^2/8)\frac1{\tau},
\ee
and
\be
E_2^{H}(\tau)=(1+\tau^2/32)\frac1{\tau}\frac{(1+\tau^2/8)}{(1+\tau^2/16)}.
\ee

\section {Optimizing the Hamiltonian energy}
\la{heop}

The Hamiltonian energy depends on both $\za_1$ and $\ka_1$ via $\ga=\sinh(u)/\ka_1$.
We derive below the {\it fundamental convergence formula for the Hamiltonian energy}.
If the thermodynamic energy is of order $\ep^{2n}$ with $u$ given by (\ref{ufin}), then
\ba
\sinh(u)&=&
\sum_{k=0}^{\infty}\frac{\ep^{2k+1}}{(2k+1)!}(1-C\ep^{2n})^{2k+1}
=\sum_{k=0}^{\infty}\frac{\ep^{2k+1}}{(2k+1)!}(1-C(2k+1)\ep^{2n}+\cdots)\nn\\
&=&\sum_{k=0}^{n-1}\frac{\ep^{2k+1}}{(2k+1)!}+\left(\frac1{(2n+1)!}-C\right)\ep^{2n+1}+\cdots.
\ea
where $C=(1-\delta_{2n+2})/(2n+2)!$.
If the algorithm is truly of $2n$ order, then as we will see below,
$\ka_1$ must be correct to $(2n-1)$ order as compared to $\sinh(\ep)$.
This is clearly true for PA and 4A.
In these cases,
\be 
\ka_1=\sum_{k=0}^{n-1}\frac{\ep^{2k+1}}{(2k+1)!}+\frac{\delta'_{2n+1}}{(2n+1)!}\ep^{2n+1}+\dots
\ee
and
\ba
\ga
&=& \frac{\sum_{k=0}^{n-1}\frac{\ep^{2k}}{(2k+1)!}+\left(\frac1{(2n+1)!}-C\right)\ep^{2n}+\cdots}
{\sum_{k=0}^{n-1}\frac{\ep^{2k}}{(2k+1)!}+\frac{\delta'_{2n+1}}{(2n+1)!}\ep^{2n}+\dots} \nn\\
&=& \frac{\sum_{k=0}^{n-1}\frac{\ep^{2k}}{(2k+1)!}+\frac{\delta'_{2n+1}}{(2n+1)!}\ep^{2n} 
+\left(\frac{1-\delta'_{2n+1}}{(2n+1)!}-C\right)\ep^{2n}+\cdots}
{\sum_{k=0}^{n-1}\frac{\ep^{2k}}{(2k+1)!}+\frac{\delta'_{2n+1}}{(2n+1)!}\ep^{2n}+\dots} \nn\\
&=& 1+ \frac{
\left(\frac{1-\delta'_{2n+1}}{(2n+1)!}-C\right)\ep^{2n}+\cdots}
{\sum_{k=0}^{n-1}\frac{\ep^{2k}}{(2k+1)!}+\frac{\delta'_{2n+1}}{(2n+1)!}\ep^{2n}+\dots} \nn\\
&=&1+D\ep^{2n}+O(\ep^{2n+2}),
\ea
where
\be
D=\left(\frac{(1-\delta^\prime_{2n+1})}{(2n+1)!}-\frac{(1-\delta_{2n+2})}{(2n+2)!}\right).
\la{dd}
\ee
The prefactor is then
\be
\rho_H=\frac1{2}(\ga+\frac1{\ga})=1+\frac12 D^2(\ep^{2n})^2+\cdots
\la{hpre}
\ee
and therefore the Hamiltonian energy convergence formula
\be
E^H_N(\tau)
\rightarrow \left (1+\frac12 D^2(\ep^{2n})^2+\cdots \right ) E(\tau).
\la{hcong}
\ee
Thus if $\ka_1$ is correct to $(2n-1)$ order as compared to $\sinh(\ep)$, then
the Hamiltonian energy will always converge from above,
at {\it twice} the order of the thermodynamic energy. 
This is due to the fact that exact
ground state $|\psi_0\rangle$ is an eigenstate of the Hamiltonian, $H|\psi_0\rangle=E_0|\psi_0\rangle$.
Given a trial state $|\phi \rangle$ approximating $|\psi_0 \rangle$
with error $|\delta\phi \rangle$,
\be
|\phi \rangle=|\psi_0 \rangle+|\delta\phi \rangle,
\ee
the Hamiltonian energy is given by
\ba
E_T=\frac{\langle \phi |H|\phi \rangle}{\langle \phi |\phi \rangle}
&=&\frac{E_0(1+\Delta)+\langle \delta\phi |H|\delta\phi \rangle}
{(1+\Delta)+\langle \delta\phi |\delta\phi \rangle}\nn\\
&=&\frac{E_0+\langle \delta\phi |H|\delta\phi \rangle/(1+\Delta)}
{1+\langle \delta\phi |\delta\phi \rangle/(1+\Delta)}=E_0+O(\delta\phi ^2),
\ea
where $\Delta=2\langle \psi_0 |\delta\phi \rangle$ is the overlap. 
Because of the eigenstate
condition, the error in the Hamiltonian energy is always quadratic 
in the error wave function $\delta\phi$. This does not hold for any other observable,
such as correlation functions, densities, kinetic and potential energy separately, or
the thermodynamic energy, for which the exact ground state is not an eigenstate. 
In these latter cases, the leading error is the overlap $\Delta$, of order $O(\delta\phi)$,
which is the order of the algorithms in PIMC.

For the PA-TI pair, $n=1$, $\delta'_3=0$, $\delta_4=2\al 4!$,
\be
D=\frac{1+16\al}{8}
\ee
and therefore 
\be
E_N^H\rightarrow\left(1+\frac12\left(\frac{1+16\al}{8}\right)^2\ep^4+\cdots\right)E(\tau).
\ee
This agrees with (\ref{hpalm}) for $\al=0$ and matches (\ref{htilm}) when $\al=1/48$.
The Hamiltonian energy error coefficient increases with $\al$, explaining why, despite TI's
fourth-order thermodynamic energy, its fourth-order Hamiltonian energy is worse than that of PA.

For the 4A-4A$^\prime$ pair, from (\ref{ka4a}) and (\ref{za4a}), $n=2$,
\be
\delta'_5=\frac{5!}{144}(1-\al)=\frac56(1-\al),\qquad \delta_6=\frac{6!}{864}(1+\al)=\frac56(1+\al),
\ee
and
\be
D=\frac{1-\frac56(1-\al)}{5!}-\frac{1-\frac56(1+\al)}{6!}=\frac{1+7\al}{864},
\ee
\be
E_N^H\rightarrow\left(1+\frac12 \Bigl(\frac{1+7\al}{864}\Bigr)^2\ep^8+O(\ep^{10})\right)E_N(\tau).
\ee
The convergence here is eight order because 4A is a truly fourth-order algorithm regardless of
the choice of $\al$. Changing $\al$ from 0 to 1/5 increases the
order of the thermodynamic energy from four to six, but does not change
the order of the Hamiltonian energy, only again increases its error coefficient. 

\begin{figure}[hbt]
	
	\includegraphics[width=0.80\linewidth]{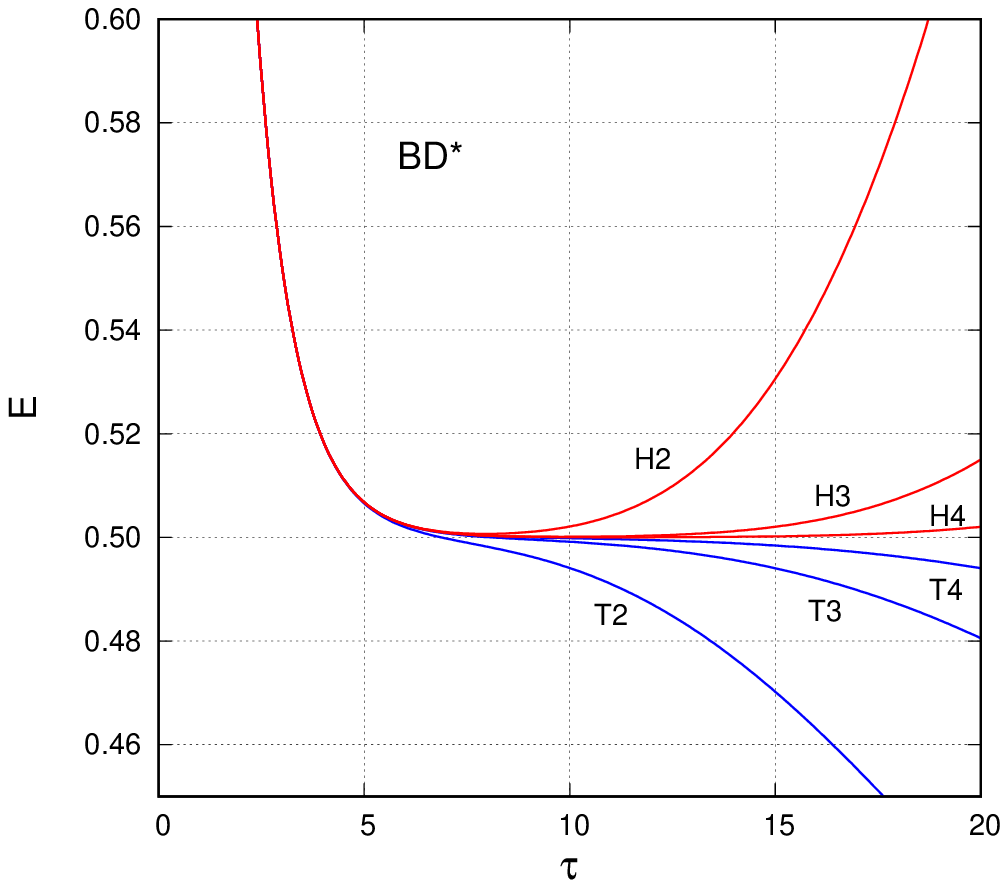}
	\caption{ (color online) 
		The sixth-order thermodynamic (T) and twelfth order Hamiltonian (H) energies at $N=2,3,4$ 
		for the algorithm BD* as a function of $\tau=N\ep$.
	}
	\la{engpb}
\end{figure}

\begin{figure}[hbt]
	\includegraphics[width=0.80\linewidth]{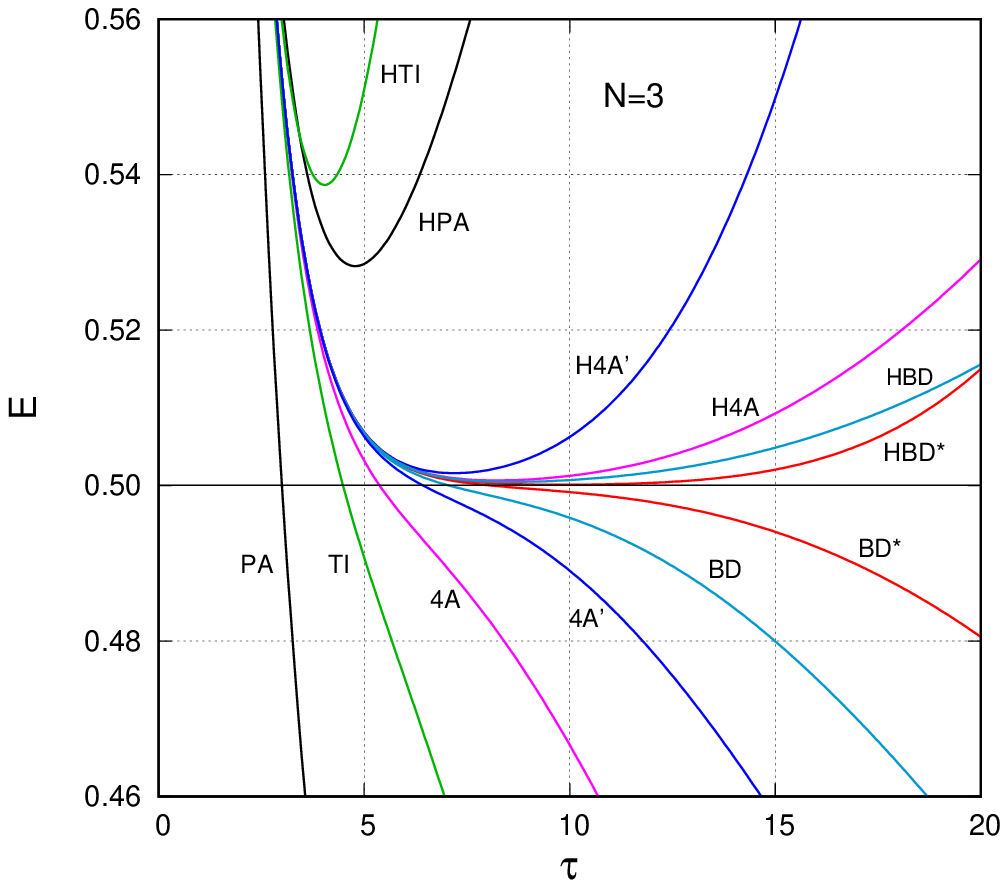}	
	\caption{ (color online)
		The $N=3$ thermodynamic and Hamiltonian energies (same color) as functions of $\tau=N\ep$
		for various short-time propagators.
	}
	\la{engn}
\end{figure}

The above discussion suggests that one should revisit the optimization of BD$'$. Instead of maximizing 
$\delta_8$ so that it is as close as possible to an eighth-order thermodynamic energy algorithm,
one should tune $t_1$ for $\delta'_5=1$ so that the Hamiltonian energy is twice that of the
sixth-order thermodynamic energy, to the {\it twelfth order}. This is achieved with
$\al=0.142872$, $t_1=0.264654$ with $n=3$, $\delta'_7=0.836636$, $\delta_8=0.987464$, giving
\be
D=3.210258\times 10^{-5},
\ee
and prefactor
\be
\rho_H=1+5.152878\times 10^{-10}\ep^{12}.
\ee
The energies of this propagator is denoted as BD$^*$ in Fig.\ref{engpb}.
Its Hamiltonian energy tabulated in Table \ref{tab2} as HDB$^*$ attained the $\tau=5$ energy
considered by Sakkos {\it et al.}\cite{sak09} at $N=3$. 
To show more clearly the energy convergence of all algorithms considered at small $N$,
we plot in Fig.{\ref{engn}}, both the thermodynamic and Hamiltonian energy of each
algorithm only for $N=3$.

Surprisingly, for Sakkos {\it et al.}'s\cite{sak09} CA algorithm, 
no real value of $t_0$ can force $\delta'_5=1$ 
and this short-time propagator cannot be optimized to yield a twelfth order Hamiltonian energy.

As can be seen in Fig.{\ref{engn}}, $\tau=5$ is still too short to reach the ground state energy. 
In Fig.\ref{congth}, we show the convergence of each algorithm's energies at $\tau=10$ as a function of 
of the step size $\ep$.

\begin{figure}[hbt]
	\includegraphics[width=0.80\linewidth]{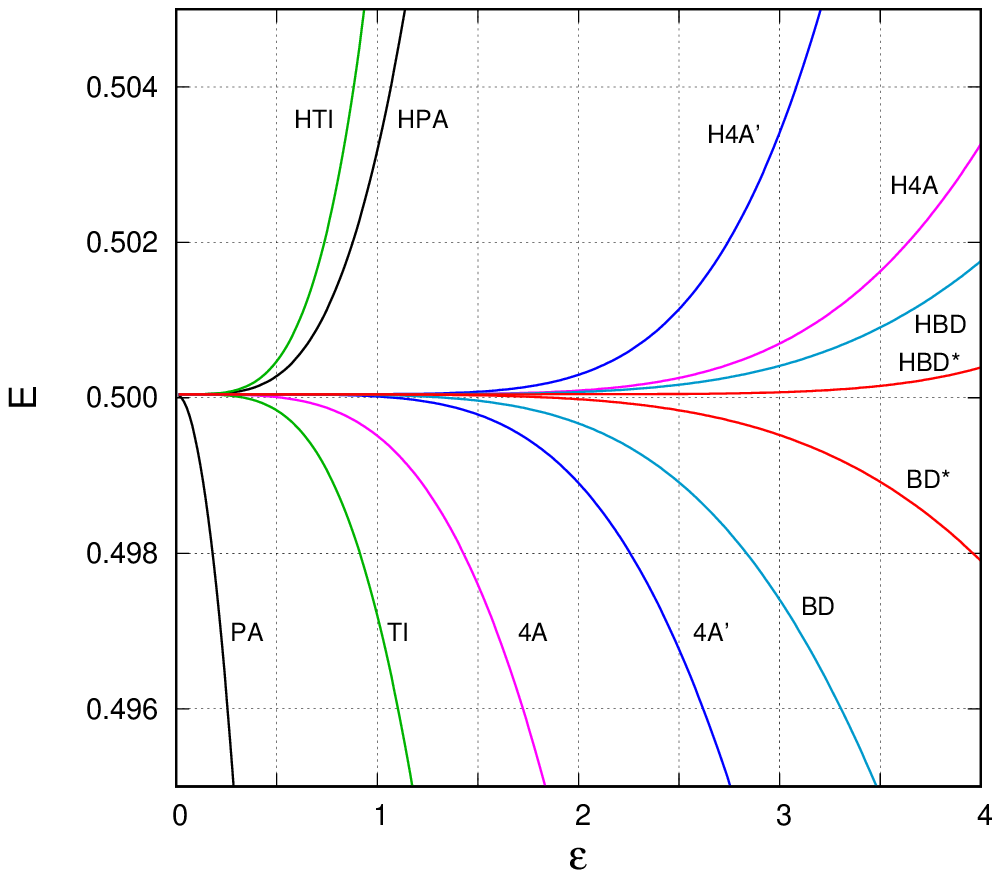}
	\caption{ (color online)
		The convergence order of thermodynamic and Hamiltonian energies (same color) for various short-time propagators
		as a function of the step size $\ep$ at a fixed $\tau=10$.
	}
	\la{congth}
\end{figure}

\section {Conclusions and future directions}
\la{con}

In examining all six algorithms in Fig.\ref{engpa}, a number of conclusions can be easily drawn. 
1) The convergence of 
PA's thermodynamic energy is exceedingly poor. The improvement in TI's thermodynamic energy is 
significant, but still modest. From Fig.\ref{congth}, both of their Hamiltonian energies, which are fourth-order,
are not much better than TI's fourth-order thermodynamic energy. 
2) There are great improvements with the use of the truly fourth-order algorithm 4A.
While its thermodynamic energy is better than that of TI, it is the convergence of its eighth-order Hamiltonian energy 
that is most dramatic. 3) With one more free-parameter, it is possible to fine 
tune the algorithm to achieve a twelfth-order convergent Hamiltonian energy, as in BD$^*$. 4) From Fig.\ref{congth},
faster convergence is obtained by computing the Hamiltonian energy as in 4A, with only one evaluation of
the double commutator, then by computing the thermodynamic energy with three evaluation of the commutator,
as in 4A$'$. Therefore, when using fourth-order algorithms, it is always more efficient to compute the Hamiltonian energy.

These conclusions can be generalized to non-harmonic and pairwise interactions. The simplest is the use algorithm 4A to compute the eighth-order Hamiltonian energy. In this case, the gradient potential $[\hV,[\hT,\hV]$ 
remains at the center and will not complicate the evaluation of the Hamiltonian energy involving
$\hV$ and $\hT$ near the edge. 
The next best thing is to fixed $\al=0$ (forcing the gradient potential away from the edge) in the BD algorithm and vary $t_1$ to minimize the Hamiltonian energy's eighth order error to zero as much as possible.

Since the harmonic oscillator Hamiltonian is separable, the analytical results of this work can be generalized to
any dimension. Because most finite systems, such as nuclei, helium droplets, quantum dots, etc., 
can be built on the basis of harmonic oscillator wave functions, this work can also 
add insights to these studies. However, an analytical model for PIMC offers the best opportunity 
to study the origin, and possibly the solution, of the sign problem. Therefore, the
most timely application of this work is the use of PIMC or PIMCGS\cite{chin20,sar00,yan17} algorithms 
with high order Hamiltonian energies to solve fermion problems in two or three dimensions.
This is because if the exact fermion propagator were known, its trace would be non-negative
and there would be no sign problem. 
The trace of two {\it approximate} propagators is also
non-negative. The sign problem begins mildly with three anti-symmetric propagators and becomes 
intractable (depending on the number of fermions) only at more than three propagators. 
As shown in this work, higher order algorithms can converge to the harmonic ground state 
with only three propagators, thereby suggesting that the problem of harmonic fermions is solvable.
Since the sign of the discrete path integral is independent of the interaction, 
one has good reason to explore further whether this high order propagator
approach can be a viable method of solving the general fermion problem without completely 
eliminating the sign problem.

\end{document}